%% file: ae07.tex
\documentclass[fleqn, 12pt]{article}
\usepackage{url}
\usepackage{amsmath}
\usepackage{amsfonts}
\usepackage{amssymb}
\usepackage{handbookbib}

\usepackage{color}
\usepackage{sgame}
\usepackage{graphics}

\usepackage{latexsym}

\include{ecmds}
\include{header}

\include{layout}

\usepackage{srcltx}
\usepackage{srctex}

\usepackage{changebar}

\title{Sequential mechanism design}

\author{Krzysztof R. Apt \\
\emph{CWI, Amsterdam, the Netherlands}, \\
\emph{University of Amsterdam} \\[2mm]
Arantza Est\'{e}vez-Fern\'{a}ndez \\
\emph{CWI}
}

\begin{document}
\maketitle

\begin{abstract}
  In the customary VCG (Vickrey-Clarke-Groves) mechanism truth-telling
  is a dominant strategy.  In this paper we study the sequential VCG
  mechanism and show that other dominant strategies may then exist. We
  illustrate how this fact can be used to minimize taxes using
  examples concerned with Clarke tax and public projects.
\end{abstract}

\section{Introduction}
\label{sec:intro}

\subsection{Motivation}

One of the basic assumptions of game theory is that each player is
rational, which means that he seeks to maximize his own utility.
However, in many circumstances it is also natural to assume that
players, when facing a choice, will seek to maximize the utilities of
other players, as well.

Such an additional assumption can be used to capture in the
game-theoretic framework a 'social attitude' of the players.  To quote
from \cite[ page 109]{Bow04} (both emphases in the text):
``\emph{Other-regarding} preferences include spite, altruism, and
caring about the relationship among the outcomes for oneself and
others. [$\dots$] \emph{The key aspect of other-regarding preferences
  is that one's evaluation of a state depends on how it is experienced
  by others.}''  Bowles also provides the following elegant quote from
Dalai Lama: ``The intelligent way to be selfish is to work for the
welfare of others''.

This additional assumption is also natural in games in which
players may or will play repeatedly. Then punishing the other players
may have an adverse effect on future rounds in which the punished
players may reciprocate. Further, the players who found out that they
were punished when another alternative existed may simply refuse to
engage in future rounds of the game.

In strategic games the attitude of other players is detected only
a posteriori. In contrast, when a game is played sequentially, the
attitude of players who already moved is known a priori to
players who still need to move.  To illustrate these matters on a
simple example consider the following strategic game: 
\II

\begin{game}{2}{3}
      & $L$   & $C$   & $R$\\
$T$   & $5,1$ & $0,0$ & $0,0$\\
$B$   & $0,0$ & $1,5$  & $5,5$
\end{game}

%\begin{game}{2}{2}
%      & $L$    & $R$\\
%$T$   &$5,1$   &$0,0$\\
%$B$   &$0,0$   &$5,5$
%\end{game}
% \II

% \II

% \begin{game}{2}{2}
%       & $L$    & $R$\\
% $T$   &$5,1$   &$0,0$\\
% $B$   &$0,0$   &$5,5$
% \end{game}
\II

\NI 
Suppose this game is played sequentially under the usual
assumption that each player is rational and that this fact is a common
knowledge. So second player always chooses a best response and first
player knows this.

If player 1 begins, then strategy $T$ guarantees him a payoff of 5, but
player 2 will only receive 1. Strategy $B$ also guarantees player 1
a payoff of 5, but only if player 2 \emph{is willing} to select then
strategy $R$.  If player 2 begins, then he will choose either $C$ or
$R$. Both strategies guarantee him a payoff of 5. However, in the first
case player 1 will receive only 1, while in the latter case he will
receive 5.  So irrespectively of the order of the play 'socially
responsible' players would select in this game strategies $B$ and
$R$.

\subsection{Sequential VCG mechanism}

In this paper we consider such matters in the context of
\emph{Vickrey-Clarke-Groves mechanism}, in short \emph{VCG mechanism}.
Its purpose is to induce players to reveal their true types
(preferences), usually in various types of auctions or matters
concerning public projects.  When this is the case, the mechanism is
called \emph{strategy-proof}. The VCG mechanism achieves this by means
of transfers. In an important special case of the VCG mechanism,
called \emph{Clarke mechanism}, the transfers become taxes imposed on
the players. The underlying game-theoretical framework is that of a
revelation-type pre-Bayesian game, see \cite{AMT06}.

In the VCG mechanism the players move simultaneously and do not know
each other types or utilities (except in degenerated situations), so
the above discussion of the attitudes towards the other players does
not apply.  However, when players move sequentially the situation
changes because each player knows the types reported by the previous
players.  In the resulting set-up the above considerations about
attitudes naturally apply. In particular, in the case of sequential
Clarke mechanism, the only way player $i$ can increase the payoff of
player $j$ is by reducing player's $j$ taxes, which leads us to an
analysis when taxes can be minimized.

Sequential VCG mechanism applies to a realistic situation
in which there is no central authority that computes and imposes
taxes and the order in which the players move depends on Nature.

In this paper we study these matters in a systematic way, by analyzing
dominant strategies in the sequential VCG mechanism and by relating
them to dominant strategies in sequential pre-Bayesian games.
Sequential VCG mechanism is relevant for various types of auctions and
for various matters concerned with public projects. The latter area is
of natural interest to us, as one is then naturally led to the problem
of maximizing social welfare through a minimization of taxes.  This
explains why our three examples are concerned with Clarke tax and
public projects.

\subsection{Related work}

To our knowledge sequential mechanism design was studied only in the context
of implementation theory. This theory focusses on a related, but different issue
of designing a game whose set of equilibria coincides with the outcomes
of a given multi-valued decision function.
An early reference on sequential implementation is \cite{MR88} in which
the implementation by means of a subgame perfect equilibrium is studied.

More recently, \cite{PS04} studied a way of realizing the VCG
mechanism with the role of the central planner reduced to a minimum.
This leads to a distributed implementation of VCG mechanism.

The problem of minimizing taxes was recently addressed in
\cite{Cav06}, who studied the issue of redistributing the taxes in the
(customary) VCG mechanism, so with truth-telling as the dominant
strategy.  Instead, in our approach we focus on minimizing taxes by
means of dominant strategies that differ from truth-telling.

The consequences of sequentiality have also been studied in voting
theory and private contributions to public goods.  In particular,
\cite{DP00} explore the relationship between simultaneous and
sequential voting games and \cite{Var94} studies the behavior of
players depending on which position they have to take a decision.

\subsection{Plan of the paper}

The paper is organized as follows.  In the next section we recall the
VCG mechanism and Clarke mechanism by focussing on the \emph{decision
  problems}.  Next, in Section \ref{sec:truthful}, we clarify
why in many natural circumstances truthful reporting in VCG
mechanism is indeed necessary.  Then, in Section \ref{sec:sequential1},
we consider sequential decision problems and in Section \ref{sec:Bayes}
we clarify the relation with sequential pre-Bayesian games.

The sequential VCG mechanism is discussed in Section \ref{sec:vcg}.
In the last three sections of the paper we consider specific instances
concerned with Clarke tax and public projects.  In each case we give
dominant strategies that minimize players' taxes and are
different from truth-telling.

\section{Preliminaries}\label{sec:prelim}

%The VCG mechanism is a revelation-type pre-Bayesian game in which for
%each player the identity function is a dominant strategy.  To define
%it and motivate it we need to introduce the last ingredient, the

We recall here briefly the VCG mechanism, see, e.g. \cite{Jac03}.
Assume a set of \oldbfe{decisions} $D$, a set $\{1, \LL, n\}$ of players, 
for each player a set of \oldbfe{types} $\Theta_i$ and a \oldbfe{utility function}
$
v_i  : D \times \Theta_i \myra \cal{R}.
$

A \oldbfe{decision rule} is a function $f: \Theta \myra D$, where
$\Theta := \Theta_1 \times \cdots \times \Theta_n$.  We call the tuple
\[
(D, \Theta_1, \LL, \Theta_n, v_1, \LL, v_n, f)
\]
a \oldbfe{decision problem}.

Given a decision problem one is interested in
the following sequence of events:

\begin{enumerate} \smallromani
\item each player $i$ receives a type $\theta_i$,

\item each player $i$ announces to the central planner a type $\theta'_i$;
this yields a joint type $\theta' := (\theta'_1, \LL, \theta'_n)$,
\label{item:2}

\item the central planner makes the decision $d := f(\theta')$,
and communicates it to each player,
\label{item:3}

\item the resulting utility for player $i$ is then
$v_i(d, \theta_i)$.
\end{enumerate}

In mechanism design one is interested in the ways of inducing the players to submit
their true types. This motivates the following two concepts.

A decision rule $f$ is called \oldbfe{strategy-proof}
if for all $\theta \in \Theta$, $i \in \C{1,\LL,n}$ and
$\theta'_i$
\[
v_i(f(\theta_i, \theta_{-i}), \theta_i) \geq
v_i(f(\theta'_i, \theta_{-i}), \theta_i).
\]
Intuitively, this means that submitting one's true type ($\theta_i$) is
better than submitting another type ($\theta_i'$).
That is, false submission does not get one better off.

A decision rule $f$ is called \oldbfe{efficient} if for all $\theta \in
\Theta$ and $d' \in D$
\[
\sum_{i = 1}^{n} v_i(f(\theta), \theta_i) \geq \sum_{i = 1}^{n} v_i(d', \theta_i).
\]
Intuitively, this means that for all $\theta \in \Theta$, $f(\theta)$
yields a decision $d$ for which the \oldbfe{society benefit}, defined
as
$
\sum_{i = 1}^{n} v_i(d, \theta_i),
$
is maximal.

%Mechanism design consists then of designing specific strategy-proof
%decision rules, for example ones that are efficient.

%\subsection{The VCG mechanism}

Recall that the VCG mechanism is constructed by combining
decision rules with transfer payments.  It is obtained by first
modifying a decision problem $(D, \Theta_1, \LL, \Theta_n, v_1,
\LL, v_n, f)$ to the following one:

\begin{itemize}

\item the set of decisions is
$
D \times \mathbb{R}^n,
$

\item the decision rule is a function
$
(f,t): \Theta \myra D \times \mathbb{R}^n,
$
where $
t: \Theta \myra \mathbb{R}^n
$
and
$
(f,t)(\theta) := (f(\theta), t(\theta)),
$

\item each utility function for player $i$ is a function $u_i: D \times \mathbb{R}^n \times \Theta_i \myra \mathbb{R}$
defined by
$
u_i(d,t_1, \LL, t_n, \theta_i) :=  v_i(d, \theta_i) + t_i.
$

\end{itemize}

So when the received type of player $i$ is $\theta_i$ and his
announced type is $\theta'_i$, his utility is $u_i((f,t)(\theta'_i,
\theta_{-i}), \theta_i) = v_i(f(\theta'_i, \theta_{-i}), \theta_i) +
t_i(\theta'_i, \theta_{-i})$, where $\theta_{-i}$ are the types
announced by the other players.

We call then $(D \times \mathbb{R}^n, \Theta_1, \LL, \Theta_n, u_1, \LL,
u_n, (f,t))$ a \oldbfe{transfer-based decision problem} and refer to
$t$ as the \oldbfe{transfer function}.

%We call the
%outcome of this modification
%a \oldbfe{decision problem with the transfers}.
%
The \oldbfe{VCG mechanism} is obtained by using the transfer function
$
t := (t_1, \LL, t_n),
$
where for all $i \in \C{1, \LL, n}$

\begin{itemize}
\item
$
h_i: \Theta_{-i} \myra \mathbb{R}
$
is an arbitrary function,

\item $t_i : \Theta \myra \mathbb{R}$ is defined by\footnote{Here and below $\sum_{j\not=i}$ is a
  shorthand for the summation over all $j \in \{1,\LL,n\}, \ j
  \not=i$.} $t_i(\theta) :=
  h_i(\theta_{-i}) + \sum_{j \neq i} v_j(f(\theta), \theta_j)$.
\end{itemize}
Intuitively, the sum $\sum_{j \neq i} v_j(f(\theta), \theta_j)$ represents the
society benefit from the decision $f(\theta)$, with player $i$ excluded.

The VCG mechanism depends on the sequence of functions $h_{1}, \LL, h_{n}$.
Occasionally we shall refer to `each' mechanism to stress that the result does not depend
on the choice of these functions.
Finally, recall the following crucial result.
\II

\NI
\textbf{VCG Theorem}
Suppose the decision rule $f$ is efficient. Then in each VCG mechanism 
the decision rule $(f,t)$ is strategy-proof.
\II

\Proof
For all $\theta \in \Theta$, $i \in \C{1, \LL, n}$ and
$\theta'_i$ we have by definition of the VCG mechanism and the fact that
$f$ is efficient:

\begin{align*}
u_i((f,t)(\theta_i, \theta_{-i}), \theta_i) &=
\sum_{j=1}^{n} v_i(f(\theta_i, \theta_{-i}), \theta_i) +  h_i(\theta_{-i})\\
&\geq \sum_{j=1}^{n} v_i(f(\theta'_i, \theta_{-i}), \theta_i) +  h_i(\theta_{-i}) \\
& = u_i((f,t)(\theta'_i, \theta_{-i}), \theta_i).
\end{align*}
\HB
\VV

%\subsection{Clarke tax}

In each VCG mechanism given the sequence $\theta$ of announced types,
$t(\theta) = (t_1(\theta), \LL, t_n(\theta))$ is the sequence of the
resulting \oldbfe{payments} that the players have to make.
If $t_i(\theta) \geq 0$, we say that player $i$
\oldbfe{receives} the payment $t_i(\theta)$ and otherwise that player
$i$ \oldbfe{makes} the payment $|t_i(\theta)|$.

%We now say that the transfer function $t$ is
%  \begin{itemize}
%  \item \oldbfe{feasible} if $\sum_{i = 1}^{n} t_i(\theta) \leq 0$ for all $\theta$,

%  \item \oldbfe{balanced} if $\sum_{i = 1}^{n} t_i(\theta) = 0$ for all $\theta$.

%  \end{itemize}

A special case of the VCG mechanism, called \oldbfe{Clarke mechanism}, is obtained by
using
\[
h_i(\theta_{-i}) := - \max_{d \in D} \sum_{j \neq i} v_j(d, \theta_j).
\]
So then
\[
t_i(\theta)  := \sum_{j \neq i} v_j(f(\theta), \theta_j) - \max_{d \in D} \sum_{j \neq i} v_j(d, \theta_j).
\]
Hence for all $\theta$ and $i \in \C{1,\LL,n}$ we have $t_i(\theta)
\leq 0$, which means that each player needs to make the payment
$|t_i(\theta)|$ that we call a \oldbfe{tax}.

\section{Truthful reporting in the VCG mechanism}\label{sec:truthful}

The following simple observation shows that for each player in each VCG
mechanism his payoff remains the same if his submitted type leads to
the same decision as his true type.

\begin{lemma} \label{lem:vcg}
Let $(D, \Theta_1, \LL, \Theta_n, v_1, \LL, v_n, f)$ be a decision
problem with efficient $f$. Let $\theta \in \Theta$ and $\theta'_i \in \Theta_i$.
In each VCG mechanism

\begin{enumerate} \smallromani
\item if
\begin{equation}
f(\theta'_i, \theta_{-i})=f(\theta_i, \theta_{-i}),
  \label{equ:theta}
\end{equation}
then
$u_i((f,t)(\theta'_i, \theta_{-i}), \theta_i) = u_i((f,t)(\theta_i, \theta_{-i}), \theta_i)$,

\item if 
$
f(\theta'_i, \theta_{-i}) \neq f(\theta_i, \theta_{-i})
$
and
\begin{equation}
  \label{equ:less}
\sum_{j=1}^{n} v_j(f(\theta'_i, \theta_{-i}), \theta_j) \neq \sum_{j=1}^{n} v_j(f(\theta_i, \theta_{-i}), \theta_j)
\end{equation}
then
$u_i((f,t)(\theta'_i, \theta_{-i}), \theta_i) < u_i((f,t)(\theta_i, \theta_{-i}), \theta_i)$.
\end{enumerate}

\end{lemma}
\Proof
By definition of the VCG mechanism we have
\[
u_i((f,t)(\theta'_i, \theta_{-i}), \theta_i)  = \sum_{j=1}^{n} v_j(f(\theta'_i, \theta_{-i}), \theta_j) + h_i(\theta_{-i})
\]
and
\[
u_i((f,t)(\theta_i, \theta_{-i}), \theta_i) = \sum_{j=1}^{n} v_j(f(\theta_i, \theta_{-i}), \theta_j)  + h_i(\theta_{-i}).
\]
This directly implies $(i)$. To prove $(ii)$ note that by
efficiency of $f$ and the second assumption of $(ii)$
\[
\sum_{j=1}^{n} v_j(f(\theta'_i, \theta_{-i}), \theta_j) < \sum_{j=1}^{n} v_j(f(\theta_i, \theta_{-i}), \theta_j).
\]
This implies $(ii)$.
\HB
\VV

Note that in the proof of $(i)$ efficiency of $f$ is not used and that in $(ii)$
the first assumption is implied by the second.

Observation $(i)$ is of limited use since in the VCG mechanism
players submit their types simultaneously and in general player
$i$ does not know the other submitted types $\theta_{-i}$. So he
has no way of ensuring that equality (\ref{equ:theta}) holds for
his submitted type $\theta'_i$ and the other submitted types $\theta_{-i}$.

In contrast, observation $(ii)$ is of use in a number of situations. Namely, 
if player $i$ knows the underlying decision problem $(D,\Theta_1, \LL, \Theta_n,
v_1, \LL, v_n, f)$, he can check for his true type $\theta_i$ and another type $\theta'_i$ whether
(\ref{equ:less}) holds for some $\theta_{-i}$.
If it does, he has to submit his true type $\theta_i$.

Now, in several natural instances of the VCG mechanism for all $i$ and
$\theta_i \neq \theta'_i$ inequality (\ref{equ:less}) holds for
some $\theta_{-i}$. In these instances players have to submit their
true types.  Examples include public project examples that we discuss in later sections.

Moreover, even if a player knows the types submitted by the other players he still may be forced to
submit his true type when he does not know their utility functions.
The following general result clarifies this claim.
We assume here that player $i$ knows $D, \Theta_1, \LL, \Theta_n, f$, knows that the decision function $f$ is efficient and
knows the submitted types $\theta_j$ for $j \neq i$, but does not know $v_j$ for $j \neq i$.

\begin{theorem}
Let $\Theta_i := \{\theta_i, \theta'_i\}$ and $\Theta_j := \{\theta_j\}$ for $j \neq i$ and
let $f: \Theta_1 \times \cdots \times \Theta_n \myra D$ be a decision function.
Suppose that\footnote{Intuitively, this condition states that the aggregate utility for player $i$ is strictly higher
for truthful reporting.}
\begin{equation}
  \label{equ:bigger}
v_i(f(\theta_i, \theta_{-i}), \theta'_i) + v_i(f(\theta'_i, \theta_{-i}), \theta_i) < v_i(f(\theta_i, \theta_{-i}), \theta_i) + v_i(f(\theta'_i, \theta_{-i}), \theta'_i).
\end{equation}
Then for some utility functions $v_j$, where $j \neq i$

\begin{itemize}
\item $f$ is efficient,

\item in each resulting VCG mechanism
$u_i((f,t)(\theta'_i, \theta_{-i}), \theta_i) < u_i((f,t)(\theta_i, \theta_{-i}), \theta_i).$

\end{itemize}
\end{theorem}

\Proof
For simplicity of notation, let 
\begin{align*}
 d &:= f(\theta_i, \theta_{-i})\\
 d'&:= f(\theta'_i, \theta_{-i})
\end{align*}
and
\begin{equation*}
q := v_i(d', \theta_i) - v_i(d, \theta_i)\mbox{.} 
\end{equation*}
Fix some $j \neq i$, take $\epsilon \in (0, v_i(d, \theta_i) + v_i(d', \theta'_i) - (v_i(d, \theta'_i) + v_i(d', \theta_i))]$, and define
\begin{align*}
v_j(d, \theta_j) &:= q + \epsilon\mbox{,} \\
v_j(d', \theta_j)&:= 0\mbox{,}\\
v_k(e, \theta_k) &:= 0\mbox{ for }k\neq i,j\mbox{ and arbitrary }e\mbox{.}
\end{align*}
Hence we have
\begin{align}
 \sum_{k=1}^{n}v_k(d,\theta_k)&=v_i(d,\theta_i)+q+\epsilon =v_i(d',\theta_i)+\epsilon\mbox{,}\label{th32eq1}\\
 \sum_{k=1}^{n}v_k(d',\theta_k)&=v_i(d',\theta_i)\mbox{,}\label{th32eq2}\\
 \sum_{k\not=i}v_k(d,\theta_k)+v_i(d,\theta'_i)& =v_i(d,\theta'_i)+q+\epsilon\mbox{,}\label{th32eq3}\\
 \sum_{k\not=i}v_k(d',\theta_k)+v_i(d',\theta'_i)& =v_i(d',\theta'_i)\mbox{,}\label{th32eq4}.
\end{align}

First, we show that given these utility functions $u_i((f,t)(\theta'_i, \theta_{-i}), \theta_i) < u_i((f,t)(\theta_i, \theta_{-i}), \theta_i)$ holds. We have
\begin{align*}
 u_i((f,t)(\theta'_i, \theta_{-i}), \theta_i) &= \sum_{k=1}^{n}v_k(d',\theta_k) + h_{i}(\theta_{-i})\\
 &= v_i(d',\theta_i)+h_{i}(\theta_{-i}) \\
 &<v_i(d',\theta_i)+\epsilon+ h_{i}(\theta_{-i})\\
 &=\sum_{k=1}^{n}v_k(d,\theta_k)+h_{i}(\theta_{-i})\\
 &=u_i((f,t)(\theta_i, \theta_{-i}), \theta_i)
\end{align*}
where the second equality follows by (\ref{th32eq2}), the inequality is a direct consequence of $\epsilon>0$, and the third equality  follows by (\ref{th32eq1}).\\

Next, we show that $f$ is efficient. Since we only have two type profiles we only have to show 
\begin{equation}\label{eff1}
 \sum_{k=1}^{n}v_k(d,\theta_k)\geq \sum_{k=1}^{n}v_k(d',\theta_k)
\end{equation}
and 
\begin{equation}\label{eff2}
 \sum_{k\not=i}v_k(d',\theta_k)+v_i(d',\theta'_i)\geq \sum_{k\not=i}v_k(d,\theta_k)+v_i(d,\theta'_i)\mbox{.}
\end{equation}

To prove (\ref{eff1}) note that
\begin{equation*}
 \sum_{k=1}^{n}v_k(d,\theta_k)=v_i(d',\theta_i)+\epsilon >v_i(d',\theta_i) =\sum_{k=1}^{n}v_k(d',\theta_k),
\end{equation*}
where the first equality follows by (\ref{th32eq1}) and the second one by (\ref{th32eq2}).

Next, to prove (\ref{eff2}) note that
\begin{align*}
 \sum_{k\not=i}v_k(d',\theta_k)+v_i(d',\theta'_i)&=v_i(d',\theta'_i)\\ &>v_i(d,\theta'_i)+v_i(d',\theta_i)-v_i(d,\theta_i)\\
 &=v_i(d,\theta'_i)+q\\
 &>v_i(d,\theta'_i)+q+\epsilon\\
 &=\sum_{k\not=i}v_k(d,\theta_k)+v_i(d,\theta'_i)
\end{align*}
where the first equality follows by (\ref{th32eq4}), the first inequality is a direct consequence of (\ref{equ:bigger}), the second equality follows by definition of $q$, and the last equality is a direct consequence of (\ref{th32eq3}).
\HB
\VV

As an aside note that asumption (\ref{equ:bigger}) in the above
theorem is necessary.

\begin{theorem}
Let $(D, \Theta_1, \LL, \Theta_n, v_1, \LL, v_n, f)$ be a decision
problem with efficient $f$.
Suppose that in some VCG mechanism for some $\theta \in \Theta$ and $\theta'_i \in \Theta_i$
\[
u_i((f,t)(\theta'_i, \theta_{-i}), \theta_i) < u_i((f,t)(\theta_i, \theta_{-i}), \theta_i).
\]
Then (\ref{equ:bigger}) holds.
\end{theorem}
\Proof
$u_i((f,t)(\theta'_i, \theta_{-i}), \theta_i) < u_i((f,t)(\theta_i, \theta_{-i}), \theta_i)$
implies
\[
\sum_{j \neq i} v_j(f(\theta'_i, \theta_{-i}), \theta_j) +v_j(f(\theta'_i, \theta_{-i}), \theta_i) < \sum_{j \neq i} v_j(f(\theta_i, \theta_{-i}), \theta_j) +v_j(f(\theta_i, \theta_{-i}), \theta_i)\mbox{.}
\]
By efficiency of $f$ we have
\[
\sum_{j \neq i} v_j(f(\theta_i, \theta_{-i}), \theta_j) + v_i(f(\theta_i, \theta_{-i}), \theta'_i) \leq \sum_{j \neq i} v_j(f(\theta'_i, \theta_{-i}), \theta_j) + v_i(f(\theta'_i, \theta_{-i}), \theta'_i).
\]
By adding up these inequalities we get
\[
v_i(f(\theta_i, \theta_{-i}), \theta'_i) + v_i(f(\theta'_i, \theta_{-i}), \theta_i) < v_i(f(\theta_i, \theta_{-i}), \theta_i) + v_i(f(\theta'_i, \theta_{-i}), \theta'_i)\mbox{.}
\]
\HB
\VV

We clarified here why in many natural circumstances truthful reporting
in VCG mechanism is necessary.  We now show that in sequential
VCG mechanism this does not need to be the case.

\section{Sequential decision problems}\label{sec:sequential1}

In the original set up of a decision problem all players announce
their types simultaneously. We now consider a modification of this
problem in which types are announced sequentially in a random order. For notational simplicity, we consider the order to be $1, \LL, n$. To capture
this type of situations, given the decision problem $(D, \Theta_1,
\LL, \Theta_n, v_1, \LL, v_n, f)$ we consider a modified sequence of
events in which events (\ref{item:2}) and  (\ref{item:3})
of Section \ref{sec:prelim}
are replaced by:
\medskip

$(ii)^{\prime}$ successively stages $1, \LL, n$ take place, where in stage $i$
  player $i$

announces to the \emph{other players} a type $\theta'_i$;

this yields a joint type $\theta' := (\theta'_1, \LL, \theta'_n)$.
\II

$(iii)^{\prime}$ each player makes the decision $d := f(\theta')$.

\medskip

We call the resulting situation a \oldbfe{sequential decision
  problem}.
So in a sequential decision problem no central planner exists and the decisions are taken by
the players themselves.
Each player $i$ \emph{knows} the types announced by players $1, \LL, i-1$.
He can then use this information to decide which type to announce.
To properly describe this situation we need to consider strategies.
In this context, a \oldbfe{strategy} of player $i$ is a function
\[
s_i: \Theta_1 \times \LL \times \Theta_i \myra \Theta_i.
\]

We then assume that in the considered
sequential decision problem each player uses a strategy $s_i(\cdot)$ to select
the type he will announce.
We say then that the strategy $s_i(\cdot)$ of player $i$ is \oldbfe{dominant}
if for all $\theta \in \Theta$, $i \in \C{1,\LL,n}$ and
$\theta'_i$
\[
v_i(f(s_i(\theta_1,\LL, \theta_i), \theta_{-i}), \theta_i) \geq
v_i(f(\theta'_i, \theta_{-i}), \theta_i).
\]
In this context, $\theta_1, \LL, \theta_{i-1}$ are the announced types of players
$1, \LL, i-1$, while $\theta_i$ is the type player $i$ has received.

Consider now the \oldbfe{projection function}
\[
\pi_i: \Theta_1 \times \LL \times \Theta_i \myra \Theta_i,
\]
where $\pi_i(\theta_1,\LL,,\theta_i) :=\theta_{i}$.
Note that $\pi_{1}(\cdot)$ is the identity function.
In this context the projection function $\pi_{i}(\cdot)$ as a strategy for
player $i$ corresponds to his truth-telling.

We have then the following observation the proof of which is immediate and omitted.

\begin{note} \label{note:equiv}
  Given a decision problem $(D, \Theta_1, \LL, \Theta_n, v_1, \LL,
  v_n, f)$ the decision rule $f$ is strategy-proof iff for all $i \in
  \C{1,\LL,n}$ the projection function $\pi_i(\cdot)$ is a dominant
  strategy for player $i$ in the corresponding sequential decision
  problem.
\end{note}

\section{Sequential pre-Bayesian games}\label{sec:Bayes}

Before we consider sequential VCG mechanisms let us clarify the connection
between sequential decision problems and strategic games.
To this end we consider a modification of
pre-Bayesian games (see e.g. \cite{AMT06}).
These games are distinguished by the fact that each player has a
private type on which he can condition his strategy.

\subsection{Pre-Bayesian games}

Recall first that a  \oldbfe{pre-Bayesian game} for $n$ players consists of

\begin{itemize}
\item a set $A_i$ of \oldbfe{actions},

\item a set  $\Theta_i$ of \oldbfe{types},

\item a \oldbfe{payoff function}
\[
p_i : A_1  \times \LL \times A_n \times \Theta_i \myra \cal{R},
\]
\end{itemize}
for each player $i$.

Let $\Theta := \Theta_1 \times \LL \times \Theta_n$
and $A := A_1 \times \LL \times A_n$.
In a pre-Bayesian game Nature moves first and provides each player
$i$ with a type $\theta_i \in \Theta_i$. Subsequently the players
simultaneously select their actions.  Each player knows only his type.
The payoff function of each player now depends on his type, so after
each player selected his action, each player knows his payoff but does
not know the payoffs of the other players.

The customary notion of a dominant
strategy is naturally adapted as follows.
First, a \oldbfe{strategy} in a pre-Bayesian game is now a function $s_i : \Theta_i
\myra A_i$.
%A \oldbfe{Nash equilibrium} of a pre-Bayesian game is
%a joint strategy $s$ such that for all
%and all $i \in \C{1,\LL,n}$ and all joint types $\theta \in \Theta$
%and actions $a_i \in A_i$
%\[
%p_i(s_i(\theta_i), s_{-i}(\theta_{-i}), \theta_i)
%\geq
%p_i(a_i, s_{-i}(\theta_{-i}), \theta_i).
%\]
Further, a strategy $s_i(\cdot)$ of player
$i$ in a pre-Bayesian game is called
 \oldbfe{dominant} if for all $a \in A$ and $\theta_i \in \Theta_i$
\[
p_i(s_i(\theta_i), a_{-i}, \theta_i) \geq p_i(a_i, a_{-i}, \theta_i).
\]

Finally, a pre-Bayesian game is of a \oldbfe{revelation-type} if $A_i
= \Theta_i$ for all $i \in \C{1,\LL,n}$. We denote then the elements
of $A_i$ by $a_i$ or $\theta_i$.  So in a revelation-type pre-Bayesian
game the strategies of a player are the functions on his set of types.

\subsection{Sequential pre-Bayesian games}

In this modification of pre-Bayesian games Nature
moves again first and provides a type $\theta_i
\in \Theta_i$ for each player $i$ and an order, say $1,\LL,n$, in
which the players sequentially select their actions. We call the resulting
game a \oldbfe{sequential pre-Bayesian game}.
In this game a \oldbfe{strategy} of player $i$ is now a function
\[
s_i: A_1 \times \LL \times A_{i-1} \times \Theta_i \myra A_i.
\]

Such a strategy $s_i(\cdot)$ of player $i$ is now called
\oldbfe{dominant} if for all $a \in A$
and $\theta_i \in \Theta_i$
\[
p_i(s_{i}(a_1, \LL, a_{i-1}, \theta_{i}), a_{-i}, \theta_i) \geq
p_i(a_i, a_{-i}, \theta_i).
\]

As before a sequential pre-Bayesian game is of a
\oldbfe{revelation-type} if $A_i = \Theta_i$ for all $i \in
\C{1,\LL,n}$.  With each decision problem $(D, \Theta_1, \LL,
\Theta_n, v_1, \LL, v_n, f)$ we can associate a revelation-type
(sequential or not) pre-Bayesian game by defining each payoff function
$p_i$ by $p_i(\theta, \theta'_i) := v_i(f(\theta), \theta'_i)$.

The following observation then clarifies the connection between
dominant strategies in the sequential decision problems and sequential
pre-Bayesian games. The proof is straightforward and therefore omitted.

\begin{note}
  A strategy $s(\cdot)$ of player $i$ is dominant in the sequential
  decision problem $(D, \Theta_1, \LL, \Theta_n, v_1, \LL, v_n, f)$
  iff it is dominant in the corresponding sequential revelation-type
  pre-Bayesian game.
\end{note}

Consequently, by Note \ref{note:equiv}, the decision rule $f$ is
strategy-proof iff for all $i \in \C{1,\LL,n}$ the projection function
$\pi_{i}(\cdot)$ is a dominant strategy for player $i$ in the
corresponding sequential revelation-type pre-Bayesian game.

%Given a sequential revelation-type pre-Bayesian game
%we now denote by $\pi_{i}$ the \oldbfe{projection function}
%\[
%\pi_i: A_1 \times \LL \times A_{i-1} \times \Theta_i \myra A_i,
%\]
%where $\pi_i(a_1,\LL,a_{i-1},\theta_i) :=\theta_{i}$.
%(Recall that $A_i = \Theta_i$.)
%Note that $\pi_{1}$ is the identity function.

%\begin{note} \label{note:equiv}
%  The decision rule $f$ is strategy-proof iff for all $i \in
%  \C{1,\LL,n}$ the projection function $\pi_i(\cdot)$ is a dominant
%  strategy for player $i$ in the corresponding sequential
%  revelation-type pre-Bayesian game.
%\end{note}

%\Proof
%\NI
%Let $\theta \in \Theta$ and $\theta'_i \in \Theta_i$.
%We have
%\[
%p_i(\pi_i(\theta_1,\LL,\theta_{i-1},\theta_i), \theta_{-i}, \theta_i) = v_i(f(\theta_i, \theta_{-i}), \theta_i)
%\]
%and
%\[
%p_i(\theta'_i, \theta_{-i}, \theta_i) = v_i(f(\theta'_i, \theta_{-i}), \theta_i).
%\]
%Then
%\[
%\mbox{$v_i(f(\theta_i, \theta_{-i}), \theta_i) \geq v_i(f(\theta'_i,
%\theta_{-i}), \theta_i)$ iff $p_i(\pi_i(\theta_1,\LL,\theta_{i-1},\theta_i), \theta_{-i}, \theta_i)
%\geq p_i(\theta'_i, \theta_{-i}, \theta_i)$},
%\]
%which concludes the proof.
%\HB

\section{Sequential VCG mechanism}\label{sec:vcg}

A particular case of sequential decision problems is the sequential
VCG mechanism. We now analyze dominant strategies in its context.  The
crucial difference between the customary set-up and the one now
considered is that player $i$ knows the types announced by players $1,
\LL, i-1$.  He can then exploit this information when choosing the
type he is to announce.  In the subsequent sections we show that in a number of
natural instances of the sequential Clarke mechanism other dominant
strategies for players exist than the projection function (that is,
truth-telling) and that they can be used to minimize taxes.

The following consequence of Lemma \ref{lem:vcg} provides us with a
simple method of determining whether a strategy is dominant in the
sequential VCG mechanism.

\begin{lemma} \label{lem:vcg1}
Let $(D, \Theta_1, \LL, \Theta_n, v_1, \LL, v_n, f)$ be a decision
problem with efficient $f$. 
In each sequential VCG mechanism,
\begin{enumerate} \smallromani
\item if for all $\theta \in \Theta$
\[f(s_i(\theta_1,\LL,\theta_i), \theta_{-i})=f(\theta_i, \theta_{-i}),\]
then strategy $s_i(\cdot)$ is dominant for player $i$,\label{lem:vcg1i}

 \item if for some ${\theta} \in \Theta$
\[f(s_i({\theta}_1,\LL,{\theta}_i), {\theta}_{-i})\not=
f({\theta}_i, {\theta}_{-i})\]
and
\[\sum_{j\in N}
v_{j}(f(s_i({\theta}_1,\LL,{\theta}_i), {\theta}_{-i}),{\theta_j})\not=
\sum_{j\in N}v_{j}(f({\theta}_i, {\theta}_{-i}),{\theta_j}),\]
then strategy $s_i(\cdot)$ is not dominant for player $i$.\label{lem:vcg1ii}
\end{enumerate}
\end{lemma}
\Proof

\NI
$(i)$
Take an arbitrary $\theta \in \Theta$ and
$\theta'_i\in\Theta_i$.  By Lemma \ref{lem:vcg}$(i)$
\[
u_i((f,t)(s_i(\theta_1,\LL, \theta_i), \theta_{-i}), \theta_i)  =
u_i((f,t)(\theta_i, \theta_{-i}), \theta_i).
\]
But by the VCG Theorem the decision rule $(f,t)$ is strategy-proof, so for every $\theta'_i\in\Theta_i$ we have
\[
u_i((f,t)(\theta_i, \theta_{-i}), \theta_i) \geq u_i((f,t)(\theta'_i, \theta_{-i}), \theta_i)\mbox{,},
\]
which concludes the proof.
\II

\NI
$(ii)$
By Lemma \ref{lem:vcg}$(ii)$ it follows
\[
u_i((f,t)(s_i(\theta_1,\LL, \theta_i), \theta_{-i}), \theta_i) < u_i((f,t)(\theta_i, \theta_{-i}), \theta_i),
\]
which concludes the proof.
% By efficiency of $f$ and the fact that
% $\sum_{j\in N}
% v_{j}(f(s_i({\theta}_1,\LL,{\theta}_i), {\theta}_{-i}),{\theta_j})\not=
% \sum_{j\in N}v_{j}(f({\theta}_i, {\theta}_{-i}),{\theta_j})$
% it follows
% $\sum_{j\in N}
% v_{j}(f(s_i({\theta}_1,\LL,{\theta}_i), {\theta}_{-i}),{\theta_j})<
% \sum_{j\in N}v_{j}(f({\theta}_i, {\theta}_{-i}),{\theta_j})$. Therefore,
% \begin{align*}
%  u_i((f,t)(s_i({\theta}_1,\LL,{\theta}_i), {\theta}_{-i}),{\theta_i})&=
% \sum_{j\in N}
% v_{j}(f(s_i({\theta}_1,\LL,{\theta}_i), {\theta}_{-i}),{\theta_j})+h_i({\theta_i})\\
% &<\sum_{j\in N}
% v_{j}(f({\theta}_i, {\theta}_{-i}),{\theta_j})+h_i({\theta_i})\\
% &=u_i((f,t)({\theta}_i, {\theta}_{-i}),{\theta_i}).
% \end{align*}
\HB
\VV

Intuitively, this lemma states that if the choice of type determined
by strategy $s_i(\cdot)$ always leads to the same decision as
truth-telling, then $s_i(\cdot)$ is dominant. Besides, if there is a
type profile for which strategy $s_i(\cdot)$ leads to a different
decision than truth-telling and to which the society is not
indifferent, then the strategy is not dominant.

In the remainder of the paper we show that in a number of natural instances of the
sequential VCG mechanism natural dominant strategies different than truth-telling exist.
All of them deal with Clarke taxes in the context of public projects.

Recall that in
Clarke mechanism, if player $i$ submits a type $\theta'_i$ and the
other submitted types are $\theta_{-i}$, each player $j$ pays
the tax $|t_j(\theta'_i, \theta_{-i})|$, where
\[
t_j(\theta'_i,\theta_{-i}) := \sum_{k \neq i,j} v_k(f(\theta'_i,\theta_{-i}), \theta_j) + v_i(f(\theta'_i,\theta_{-i}), \theta'_i) -
\max_{d \in D} (\sum_{k \neq i,j} v_k(d, \theta_j) + v_i(d, \theta'_i)). 
\]

In the sequential Clarke mechanism we have $t_j(\theta'_i,\theta_{-i}) \leq 0$, so player $i$, when using
Lemma \ref{lem:vcg1}$(i)$ to minimize player's $j$ tax, solves the
following maximization problem:
\begin{equation}\label{eqclarke}
\mbox{maximize $t_j(\theta'_i, \theta_{-i})$ subject to $\theta'_i \in \Theta_i$ and $f(\theta'_i, \theta_{-i}) = f(\theta_i, \theta_{-i})$.}
\end{equation}

\section{Example: public project I}\label{sec:example1}

This example corresponds to the decision problem
$
(D, \Theta_1, \LL, \Theta_n, v_1, \LL, v_n, f),
$
where

\begin{itemize}
\item $D = \{0, 1\}$
(reflecting whether a project is cancelled or takes place),

\item each $\Theta_i$ is a set of non-negative reals, including 0 and $c$,

\item $v_i(d, \theta_i) := d (\theta_i - \frac{c}{n})$,

\item $
        f(\theta) :=
        \left\{
        \begin{array}{l@{\extracolsep{3mm}}l}
        1    & \mathrm{if}\  \sum_{i = 1}^{n} \theta_i \geq c \\
        0       & \mathrm{otherwise}
        \end{array}
        \right.
$
\end{itemize}

In this setting $c$ is the cost of the project, $\frac{c}{n}$ is the cost
share of the project for each player, and $\theta_i$ is the value
of the project for player $i$. Besides, note that the decision rule $f$ is efficient since
$\sum_{i = 1}^{n} v_i(d, \theta_i) = d(\sum_{i = 1}^{n} \theta_i - c)$.

We have then the following result.

\begin{theorem} \label{thm:dom1}
The following strategy is dominant for player $i$
in the sequential Clarke mechanism for the above decision problem:
\[
s_i(\theta_1, \LL, \theta_i) :=
 \left\{
        \begin{array}{l@{\extracolsep{3mm}}l}
        \theta_i    & \mathrm{if}\  \sum_{j=1}^{i}\theta_j < c\mathrm{\ and \ } i<n\mbox{,} \\
        0  & \mathrm{if}\  \sum_{j=1}^{i}\theta_j < c\mathrm{\ and \ } i=n\mbox{,} \\
        c       & \mathrm{if \ }\ \sum_{j=1}^{i}\theta_j \geq c\mbox{.}
        \end{array}
        \right.
\]
\end{theorem}
\Proof 
By Lemma \ref{lem:vcg1} it suffices to show
that $f(s_i(\theta_1,\LL,\theta_i), \theta_{-i})=f(\theta_i, \theta_{-i})$. For this we consider three cases.
\II

\NI
\emph{Case 1} $s_i(\theta_1,\LL,\theta_i)=\theta_{i}$.

Then $f(s_i(\theta_1,\LL,\theta_i), \theta_{-i})=f(\theta_i, \theta_{-i})$.
\II

\NI
\emph{Case 2} $s_i(\theta_1,\LL,\theta_i) = 0$.

By definition of $s_i(\cdot)$ we have $i=n$
and $c>\sum_{j=1}^{n}\theta_j \geq s_i(\theta_1,\LL,\theta_i)+\sum_{i\not=j} \theta_j$
and therefore $f(s_i(\theta_1,\LL,\theta_i), \theta_{-i})=f(\theta_i, \theta_{-i})$,
as both sides equal 0.
\II

\NI
\emph{Case 3} $s_i(\theta_1,\LL,\theta_i) = c$.

By definition of $s_i(\cdot)$ we have both
$\sum_{j=1}^{n}\theta_j \geq \sum_{j=1}^{i}\theta_j \geq c$ and
$s_i(\theta_1,\LL,\theta_i)+\sum_{i\not=j} \theta_j \geq c$
and therefore $f(s_i(\theta_1,\LL,\theta_i), \theta_{-i})=f(\theta_i, \theta_{-i})$,
as both sides equal 1.
\HB
\VV

We now prove that in the sequential Clarke mechanism considered in Theorem \ref{thm:dom1}
the strategy  $s_i(\cdot)$ of player $i$ simultaneously solves the above maximization problems
(\ref{eqclarke}) for $j \neq i$, i.e., this strategy of player $i$ minimizes the tax of every other player.
More precisely, we establish the following result.

\begin{theorem} \label{thm:dom1-taxes}
Consider the sequential Clarke mechanism
of Theorem \ref{thm:dom1} and the strategy $s_i(\cdot)$ of player $i$ introduced there.
Suppose that $s_i(\theta_1,\LL,\theta_i) \neq \theta_i$.
Then
\[
t_j(s_i(\theta_1,\LL,\theta_i),\theta_{-i}) \geq  t_j(\theta'_i, \theta_{-i})
\]
for all $j \neq i$, $\theta_{i+1} \in \Theta_{i+1}, \LL, \theta_{n} \in \Theta_n$, 
and $\theta'_i \in \Theta_i$ such that
$f(\theta'_i, \theta_{-i}) = f(\theta_i, \theta_{-i})$.

\end{theorem}

In other words, if strategy $s_i(\cdot)$ of player $i$ deviates from
truth-telling, then player $i$ minimizes taxes of other players under
the assumption that he submits a type that will not alter the decision
taken in case of truth telling by all players.

\NI
\Proof
Let
\[
g_j(\theta'_i, \theta_{-i}) := \sum_{k\not=i,j}(\theta_k-\frac{c}{n}) + \theta'_i - \frac{c}{n}.
\]
We have for all $\theta \in \Theta$, $j \neq i$ and $\theta'_i$
\[
t_j(\theta'_i, \theta_{-i}) =
 \left\{
        \begin{array}{l@{\extracolsep{3mm}}l}
        -\max_{d \in \{0,1\}}d \cdot g_j(\theta'_i, \theta_{-i})
 & \mathrm{if}\  \sum_{k \neq i}\theta_k + \theta'_i < c \\
  g_j(\theta'_i, \theta_{-i}) -\max_{d \in \{0,1\}}d \cdot g_j(\theta'_i, \theta_{-i})
& \mathrm{otherwise}
        \end{array}
        \right.
\]
\II

\NI
\emph{Case 1} $s_i(\theta_1,\LL,\theta_i) = 0$.

By definition of $s_i(\cdot)$ we have $i=n$
and $\sum_{j=1}^{n}\theta_j < c$. So $\sum_{j \neq i}\theta_j + s_i(\theta_1,\LL,\theta_i) < c$.
Also
$\sum_{j \neq i}\theta_j + \theta'_i < c$ since
$f(\theta'_i, \theta_{-i}) = f(\theta_i, \theta_{-i})$.
Hence

\[
t_j(s_i(\theta_1,\LL,\theta_i), \theta_{-i}) = -\max_{d \in \{0,1\}}d \cdot g_j(s_i(\theta_1,\LL,\theta_i), \theta_{-i})
\]
and
\[
t_j(\theta'_i, \theta_{-i}) = -\max_{d \in \{0,1\}}d \cdot g_j(\theta'_i, \theta_{-i}).
\]

But $s_i(\theta_1,\LL,\theta_i)=0 \leq \theta'_i$, so
\begin{align*}
 g_j(s_i(\theta_1,\LL,\theta_i), \theta_{-i}) &= \sum_{k\not=i,j}\theta_k + s_i(\theta_1,\LL,\theta_i) - \frac{n-1}{n} c \\
 &\leq \sum_{k\not=i,j}\theta_k + \theta'_i - \frac{n-1}{n} c \\
 &= g_j(\theta'_i, \theta_{-i}),
\end{align*}
which implies the claimed inequality.

\II

\NI
\emph{Case 2} $s_i(\theta_1,\LL,\theta_i) = c$.

Then $\sum_{k=1}^{i}\theta_k \geq c$ and hence
$\sum_{k \neq i}\theta_k + s_i(\theta_1,\LL,\theta_i) \geq c$, so $g_j(s_i(\theta_1,\LL,\theta_i), \theta_{-i})\geq 0$ and
\[
t_j(s_i(\theta_1,\LL,\theta_i), \theta_{-i}) =  g_j(s_i(\theta_1,\LL,\theta_i), \theta_{-i}) -\max_{d \in \{0,1\}}d \cdot g_j(s_i(\theta_1,\LL, \theta_i), \theta_{-i})=0.
\]

Moreover, $\sum_{k=1}^{n}\theta_n \geq \sum_{k=1}^{i}\theta_k \geq c$,
so also $\sum_{k \neq i}\theta_k + \theta'_i \geq c$ since
$f(\theta'_i, \theta_{-i}) = f(\theta_i, \theta_{-i})$.
Hence
$
t_j(\theta'_i, \theta_{-i}) =  g_j(\theta'_i, \theta_{-i}) -\max_{d \in \{0,1\}}d \cdot g_j(\theta'_i, \theta_{-i}) \leq 0.
$
\HB
\VV

Let us illustrate the above two theorems by two examples.

\begin{example} \label{exa:1}
  
  Suppose there are three players, A, B, and C whose types (values) are
  respectively 60, 70, and 250, each player can submit an arbitrary
  non-negative value, and the total cost $c$ of the project equals
  300.  In the customary situation, when the players submit their
  values simultaneously the project takes place (the efficient
  decision is 1) and we get the situation summarized in Table
  \ref{tab:1}, where $\mathbb{R}_+$ denotes the set of non-negative reals.

\begin{table}[htbp]

\begin{center}
\begin{tabular}{|c|c|c|c|c|c|}
\hline
player & value & set of types ($\Theta_i$) & Clarke tax & cost share & utility ($u_i$) \\\hline
A & $60$  & $\mathbb{R}_+$ & $0$ & $100$ & $-40$\\\hline
B & $70$  & $\mathbb{R}_+$ & $0$ & $100$ & $-30$\\\hline
C & $250$ & $\mathbb{R}_+$ & $70$ & $100$ & $150$\\\hline
\end{tabular}\vspace{0.25cm}
\end{center}
  \caption{Clarke taxes: the project takes place}
  \label{tab:1}
\end{table}

Consider now the situation in which the players submit their values sequentially
and each of them follows strategy  $s_{i}(\cdot)$.
There are three possible cases. 
The resulting taxes are summarized in Table \ref{tab:1a}.

\begin{itemize}
\item Player A is the last player. 
  
\NI
According to strategy $s_{i}(\cdot)$ players B and C will submit
their true values, since for each of them the first alternative in the
definition of $s_{i}(\cdot)$ holds.  However, player A will submit 300
since for him the third alternative holds. The tax of player B remains
0, but the tax of player C gets modified and, in accordance with the proof
of Theorem \ref{thm:dom1-taxes} (\emph{Case 2}), becomes 0.

\item Player B is the last player. 
  
\NI
The situation is analogous to the previous case.  Player B will submit
300. As a result all taxes become 0.

 \item Player C is the last player. 
   
\NI
Here players A and B will submit their true values, but player C will
submit 300. This does not modify the taxes of players A and B (which remain 0) and
player's C tax also remains 70.

\end{itemize}

\begin{table}[htbp]

\begin{center}
\begin{tabular}{|c|c|c|c|}
\hline
ordering &  $t_A$  &  $t_B$  & $t_C$ \\\hline
A B C    & $0$& $0$ & $70$  \\\hline
A C B    & $0$& $0$ & $0$  \\\hline
B A C    & $0$& $0$ & $70$  \\\hline
B C A    & $0$& $0$ & $0$  \\\hline
C A B    & $0$& $0$ & $0$  \\\hline
C B A    & $0$& $0$ & $0$  \\\hline
\end{tabular}\vspace{0.25cm}
\end{center}
  \caption{Clarke taxes in the sequential cases}
  \label{tab:1a}
\end{table}

Note that in the first two cases, according to strategy
$s_{i}(\cdot)$, if player C is second he will submit 300, but this is
irrelevant for the analysis.  We conclude that if each player follows
strategy $s_i(\cdot)$, in four out of six orderings all taxes are
reduced to 0.

This example also shows that if strategy $s_i(\cdot)$ of player $i$ does not deviate from
truth-telling, then player $i$ does not need to minimize taxes of other players.
Indeed, if player C is the last player, then according to the $s_i(\cdot)$ strategy
the second player will submit his true value, whereas submitting 300, 
a value that would not alter the decision taken,  would reduce the tax
of player C to 0. 
The problem is of course that the second player does not know which decision will be taken and hence,
by Lemma \ref{lem:vcg1}$(i)$, is bound to submit his true value.
\HB
\end{example}

In general, if $i$ is the first player for which $\sum_{j=1}^{i}\theta_j \geq c$, then according to 
strategy $s_i(\cdot)$ he will submit $c$. This reduces the taxes of all players except him to 0
(\emph{Case 2} in the proof of Theorem \ref{thm:dom1-taxes}).
Player's $i$ tax may or may not become 0. If he is not the last player, then all players
$i+1, \LL, n$ following him will also submit $c$, which will ensure that all taxes
\emph{including} the one of player $i$ will become 0.

If for no $i$, $\sum_{j=1}^{i}\theta_j \geq c$, the situation changes,
as the following example illustrates.

\begin{example} \label{exa:2}
  We change the setting of the previous example and assume that the values of players
  A, B, and C are respectively 60, 70, and 150 while the project cost remains 300. 
Now when they submit their
  values simultaneously the project does not take place (the efficient
  decision is 0) and we get the situation summarized in Table
  \ref{tab:2}.

\begin{table}[htbp]

\begin{center}
\begin{tabular}{|c|c|c|c|c|c|}
\hline
player & value & set of types ($\Theta_i$) & Clarke tax & cost share & utility ($u_i$) \\\hline
A & $60$  & $\mathbb{R}_+$ & $20$ & $0$ & $-20$\\\hline
B & $70$  & $\mathbb{R}_+$ & $10$ & $0$ & $-10$\\\hline
C & $150$ & $\mathbb{R}_+$ & $0$ & $0$ & $0$\\\hline
\end{tabular}\vspace{0.25cm}
\end{center}
  \caption{Clarke taxes: the project does not take place}
  \label{tab:2}
\end{table}

In the sequential case, according to the strategy $s_i(\cdot)$, the
first two players will submit their true types and the last player
will submit 0, since for him the second alternative in the definition
of $s_{i}(\cdot)$ holds.

Again we have three cases. The resulting taxes are summarized in Table \ref{tab:2a}.

\begin{itemize}
\item Player A is the last player. 

By submitting 0 player A reduces the tax of player B to 0, the tax of player C remains 0
and the tax of player A remains 20.

\item Player B is the last player. 

By submitting 0 player B reduces the tax of player A to 0, the tax of player C remains 0
and the tax of player B remains 10.

\item Player C is the last player. 

By submitting 0 player C reduces the taxes of players A and B to 0 and his tax remains 0.

\end{itemize}

\begin{table}[htbp]

\begin{center}
\begin{tabular}{|c|c|c|c|}
\hline
ordering &  $t_A$  &  $t_B$  & $t_C$ \\\hline
A B C    & $0$& $0$ & $0$  \\\hline
A C B    & $0$& $10$ & $0$  \\\hline
B A C    & $0$& $0$ & $0$  \\\hline
B C A    & $20$& $0$ & $0$  \\\hline
C A B    & $0$& $10$ & $0$  \\\hline
C B A    & $20$& $0$ & $0$  \\\hline
\end{tabular}\vspace{0.25cm}
\end{center}
  \caption{Clarke taxes in the sequential cases}
  \label{tab:2a}
\end{table}

So we see that in each ordering some tax gets reduced and in two out of six orderings all taxes
get reduced to 0.
\HB
\end{example}

\section{Example: public project II}\label{sec:example2}

Consider now a modification of the above example in which
each $\Theta_i$ is a real interval $[0, r_i]$, where $r_i \geq 0$.
The following is a counterpart of Theorem \ref{thm:dom1}.

\begin{theorem} \label{thm:dom2}
The following strategy is dominant for player $i$
in the corresponding sequential Clarke mechanism:

\[
s_i(\theta_1, \LL, \theta_i) :=
 \left\{
        \begin{array}{l@{\extracolsep{3mm}}l}
        \theta_i    & \mathrm{if}\  \sum_{j=1}^{i}\theta_j<c\mbox{,
}\sum_{j=1}^{i}\theta_j + \sum_{j=i+1}^{n}r_j \geq c\mathrm{, } \\
        0  & \mathrm{if}\  \sum_{j=1}^{i}\theta_j + \sum_{j=i+1}^{n}r_j  <
c\mathrm{, }\\
        r_i       & \mathrm{if \ }\ \sum_{j=1}^{i}\theta_j \geq c\mbox{.}
        \end{array}
        \right.
\]
\end{theorem}
\Proof
As in the proof of Theorem \ref{thm:dom1}, it suffices to show by
Lemma \ref{lem:vcg} that $f(s_i(\theta_1,\LL,\theta_i),
\theta_{-i})=f(\theta_i, \theta_{-i})$. For this we consider three
cases.
\II

\NI
\emph{Case 1} $s_i(\theta_1,\LL,\theta_i)=\theta_{i}$.

Then $f(s_i(\theta_1,\LL,\theta_i), \theta_{-i})=f(\theta_i, \theta_{-i})$.
\II

\NI
\emph{Case 2} $s_i(\theta_1,\LL,\theta_i)=0$.

By definition of $s_i(\cdot)$ we have $c>\sum_{j=1}^{i}\theta_j+\sum_{j=i+1}^{n}r_j
\geq \sum_{j=1}^{i-1}\theta_j+\sum_{j=i+1}^{n}r_j\geq \sum_{j\not=i}\theta_j=s_i(\theta_1,\LL,\theta_i)+\sum_{i\not=j} \theta_j$
and therefore $f(s_i(\theta_1,\LL,\theta_i), \theta_{-i})=f(\theta_i, \theta_{-i})$
as both sides equal 0.
\II

\NI \emph{Case 3} $s_i(\theta_1,\LL,\theta_i)=r_i$. By definition of
$s_i(\cdot)$ we have both $c \leq
\sum_{j=1}^{i}\theta_j\leq\sum_{j=1}^{n}\theta_j$ and $ c \leq
\sum_{j=1}^{i}\theta_j \leq \sum_{j=1}^{i-1}\theta_j + r_i \leq
s_i(\theta_1,\LL,\theta_i)+\sum_{i\not=j} \theta_j$ and therefore
$f(s_i(\theta_1,\LL,\theta_i), \theta_{-i})=f(\theta_i,
\theta_{-i})$ as both sides equal 1. 
\HB
\VV

Next, we have the following counterpart of Theorem \ref{thm:dom1-taxes}.

\begin{theorem} \label{thm:dom2-taxes}
Consider the sequential Clarke mechanism of Theorem \ref{thm:dom2}
and the strategy $s_i(\cdot)$ of player $i$ introduced there.
Suppose that $s_i(\theta_1,\LL,\theta_i) \neq \theta_i$. Then
\[
t_j(s_i(\theta_1,\LL,\theta_i),\theta_{-i}) \geq  t_j(\theta'_i,
\theta_{-i})
\]
for all $j \neq i$, $\theta_{i+1} \in \Theta_{i+1}, \LL, \theta_{n} \in \Theta_n$, 
and $\theta'_i \in \Theta_i$ such that
$f(\theta'_i, \theta_{-i}) = f(\theta_i, \theta_{-i})$.

\end{theorem}
\Proof 
The proof is analogous to that of Theorem \ref{thm:dom1-taxes}.
We consider two cases.

% Let
% \[
% g_j(\theta'_i, \theta_{-i}) :=
% \sum_{k\not=i,j}(\theta_k-\frac{c}{n}) + \theta'_i - \frac{c}{n}.
% \]
% We have for all $\theta \in \Theta$, $i \neq j$ and $\theta'_i$
% \[
% t_j(\theta'_i, \theta_{-i}) =
%  \left\{
%         \begin{array}{l@{\extracolsep{3mm}}l}
%         -\max_{d \in \{0,1\}}d \cdot g_j(\theta'_i, \theta_{-i})
%  & \mathrm{if}\  \sum_{k \neq i}\theta_k + \theta'_i < c \\
%   g_j(\theta'_i, \theta_{-i}) -\max_{d \in \{0,1\}}d \cdot g_j(\theta'_i, \theta_{-i})
% & \mathrm{otherwise}
%         \end{array}
%         \right.
% \]
\II

\NI \emph{Case 1} $s_i(\theta_1,\LL,\theta_i) = 0$.

By definition of $s_i(\cdot)$ we have
$\sum_{j=1}^{i}\theta_j+\sum_{j=i+1}^{n}r_{j}< c$. So $\sum_{j \neq
i}\theta_j + s_i(\theta_1,\LL,\theta_i)=\sum_{j \neq i}\theta_j\leq
\sum_{j=1}^{i-1}\theta_j+\sum_{j=i+1}^{n}r_{j}< c$. Also $\sum_{j
\neq i}\theta_j + \theta'_i < c$ since $f(\theta'_i, \theta_{-i}) =
f(\theta_i, \theta_{-i})$. 
\II

\NI 
\emph{Case 2} $s_i(\theta_1,\LL,\theta_i) = r_{i}$.

Then $\sum_{k=1}^{i}\theta_k \geq c$ and hence $\sum_{k \neq
i}\theta_k + s_i(\theta_1,\LL,\theta_i)=\sum_{k \neq i}\theta_k +
r_i\geq \sum_{k=1}^{i}\theta_k \geq c$.
Also $\sum_{j \neq i}\theta_j + \theta'_i \geq c$ since
$f(\theta'_i, \theta_{-i}) = f(\theta_i, \theta_{-i})$. 
\II

The rest of the proof of both cases is the same as in the proof
of Theorem \ref{thm:dom1-taxes}.
\HB
\VV

Let us illustrate now the above two theorems with the following example.

\begin{example} \label{exa:3}
  
  We modify Example \ref{exa:2} by restricting the set of types for
  each player.  The situation is summarized in Table \ref{tab:3}.  So
  when the players submit their values simultaneously, no change
  arises.

\begin{table}[htbp]

\begin{center}
\begin{tabular}{|c|c|c|c|c|c|}
\hline
player & value & set of types ($\Theta_i$) & Clarke tax & cost share & utility ($u_i$) \\\hline
A & $60$  & $[0,100]$ & $20$ & $0$ & $-20$\\\hline
B & $70$  & $[0,80]$  & $10$ & $0$ & $-10$\\\hline
C & $150$ & $[0,150]$ & $0$  & $0$ & $  0$\\\hline
\end{tabular}\vspace{0.25cm}
\end{center}
  \caption{Clarke taxes: the project does not take place}
  \label{tab:3}
\end{table}

However, in the sequential case a new situation arises when player B
is the last player.  The reason is that now the second player knows
that the project will not take place, that is for him the second
alternative in the definition of $s_{i}(\cdot)$ holds.  So the second
player will submit 0. Also player B will submit 0. As a result all
taxes will be reduced to 0.

In the other ordering of the players the situation will remain as in
Example \ref{exa:2}.  In particular, when player A is the last player
his tax will remain 20. The reason is that
the second player does not know yet that the project will not take
place, that is for him the first alternative in the definition of
$s_{i}(\cdot)$ holds.  So the second player will submit his true
value.

The situation is summarized in Table \ref{tab:3a}.

\begin{table}[htbp]

\begin{center}
\begin{tabular}{|c|c|c|c|}
\hline
ordering &  $t_A$  &  $t_B$  & $t_C$ \\\hline
A B C    & $0$& $0$ & $0$  \\\hline
A C B    & $0$& $0$ & $0$  \\\hline
B A C    & $0$& $0$ & $0$  \\\hline
B C A    & $20$& $0$ & $0$  \\\hline
C A B    & $0$& $0$ & $0$  \\\hline
C B A    & $20$& $0$ & $0$  \\\hline
\end{tabular}\vspace{0.25cm}
\end{center}
  \caption{Clarke taxes in the sequential cases, with limited sets of types}
  \label{tab:3a}
\end{table}

\end{example}

\section{Example: choosing a project}\label{sec:example3}

This example corresponds to the decision problem
$
(D, \Theta_1, \LL, \Theta_n, v_1, \LL, v_n, f),
$
where

\begin{itemize}
\item $D = \{1,\LL,m\}$ (reflecting which project takes place),

\item $\Theta_i \sse \mathbb{R}^{m}_{+}$, where for every $\theta_i\in\Theta_i$ we have $\theta_{ik}\in[0,r_{ik}]$ for $k\in\{1,\LL,m\}$,

\item $v_i(d, \theta_i) := \theta_{id}$,

\item $f(\theta) :=\arg\max\{\sum_{i=1}^{n}\theta_{ik}|k\in\{1, \LL, m\}\}$.
\end{itemize}

So each player $i$ submits a vector of $m$ non-negative
reals, reflecting his appreciation for the individual projects. Each
real $\theta_{ik}$ is player's $i$ appreciation for project $k$ and is
taken from the interval $[0,r_{ik}]$. When project $d$ is selected and
player's $i$ true type is $\theta_i$, his utility is $\theta_{id}$.  The
decision function $f$ selects the project with the largest aggregated
appreciation (with ties randomly broken).  It is easy to see that $f$ is efficient.

When players submit their types sequentially, 
player $i$ knows the submitted types $\theta_j$ for $j \in \{1, \LL, i-1\}$ and his own type
$\theta_i$. Then
$
\sum_{j=1}^{i}\theta_{j\bar{l}}\leq \sum_{j=1}^{i}\theta_{jl} + \sum_{j=i+1}^{n}r_{jl}
$ 
is the maximum possible aggregated appreciation for project $l$, as perceived by player $i$.
We use it in the following result.

\begin{theorem} \label{thm:dom3}
The following strategy is dominant for player $i$
in the sequential Clarke mechanism for the above decision problem:
\[
s_i(\theta_1, \LL, \theta_i) :=
 \left\{
        \begin{array}{l@{\extracolsep{3mm}}l}
        \theta_i    & \mathrm{if}\  \sum_{j=1}^{i}\theta_{j\bar{l}}\leq \sum_{j=1}^{i}\theta_{jl} + \sum_{j=i+1}^{n}r_{jl} \\
                    & \mbox{for all } l,\bar{l} \in\{1,\LL m\}\mbox{, }l\not=\bar{l},\\
        \textbf{0}_{i \rightarrow r_{i\bar{l}}}   & \mathrm{if}\  \sum_{j=1}^{i}\theta_{j\bar{l}} > \sum_{j=1}^{i}\theta_{jl} + \sum_{j=i+1}^{n}r_{jl} \\
                    & \mbox{for all } l\in\{1,\LL m\} \setminus \{\bar{l}\},
%        r_{i\bar{l}}e^{\{\bar{l}\}}    & \mathrm{if}\  \sum_{j=1}^{i}\theta_{j\bar{l}} > \sum_{j=1}^{i}\theta_{jl} + \sum_{j=i+1}^{n}r_{jl} \\ \mbox{for every } l\in\{1,\LL m\}\mbox{, }l\not=\bar{l},
        \end{array}
        \right.
\]
where $\textbf{0}_{i \rightarrow r_{i\bar{l}}}$ is a vector of $m$ reals in which exactly one entry, the $i$th one, is non-zero.
This entry equals $r_{i\bar{l}}$, the maximum appreciation of player $i$ for project $\bar{l}$.

%$e^{\{\bar{l}\}}\in\mathbb{R}^{m}$ with $e^{\{\bar{l}\}}_{l}=1$ if $l=\bar{l}$ and $e^{\{\bar{l}\}}_{l}=0$ otherwise.
\end{theorem}

\Proof
As in the proof of Theorem \ref{thm:dom1}, it suffices to show by
Lemma \ref{lem:vcg} that $f(s_i(\theta_1,\LL,\theta_i),
\theta_{-i})=f(\theta_i, \theta_{-i})$. For this we consider two
cases.
\II

\NI
\emph{Case 1} $s_i(\theta_1,\LL,\theta_i)=\theta_{i}$.

Then $f(s_i(\theta_1,\LL,\theta_i), \theta_{-i})=f(\theta_i, \theta_{-i})$.
\II

\NI
\emph{Case 2} $s_i(\theta_1,\LL,\theta_i)= \textbf{0}_{i \rightarrow r_{i\bar{l}}}$.

By definition of $s_i(\cdot)$ we have 
$$
\sum_{j=1}^{n}\theta_{j\bar{l}} > \sum_{j=1}^{i}\theta_{jl} + \sum_{j=i+1}^{n}r_{jl}+\sum_{j=i+1}^{n}\theta_{j\bar{l}}\geq\sum_{j=1}^{n}\theta_{jl} $$ 
for every $l\not=\bar{l}$.
Then $f(\theta_1,\LL,\theta_n)=\bar{l}$ and 
\begin{align*}
 \sum_{j\not=i}\theta_{j\bar{l}}+s_{i\bar{l}}(\theta_{1},\LL,\theta_{i})&=
 \sum_{j\not=i}\theta_{j\bar{l}}+r_{i\bar{l}}
 \geq \sum_{j=1}^{i}\theta_{j\bar{l}} \\
 &>\sum_{j=1}^{i}\theta_{jl} + \sum_{j=i+1}^{n}r_{jl}
 \geq \sum_{j=1}^{n}\theta_{jl}
 \geq \sum_{j\not=i}\theta_{jl}\\
 &=\sum_{j\not=i}\theta_{jl}+s_{il}(\theta_{1},\LL,\theta_{i})
\end{align*}
for every $l\not=\bar{l}$. Therefore, $f(s_i(\theta_1,\LL,\theta_i), \theta_{-i}) = \bar{l} = f(\theta_i, \theta_{-i})$.
\HB
\VV

Also, as in the earlier two examples strategy $s_i(\cdot)$ minimizes taxes.

\begin{theorem}
Consider the sequential Clarke mechanism of Theorem \ref{thm:dom3}
and the strategy $s_i(\cdot)$ of player $i$ introduced there.
Suppose that $s_i(\theta_1,\LL,\theta_i) \neq \theta_i$. Then
\[
t_j(s_i(\theta_1,\LL,\theta_i),\theta_{-i}) \geq  t_j(\theta'_i,
\theta_{-i})
\]
for all $j \neq i$, $\theta_{i+1} \in \Theta_{i+1}, \LL, \theta_{n} \in \Theta_n$, 
and $\theta'_i \in \Theta_i$ such that
$f(\theta'_i, \theta_{-i}) = f(\theta_i, \theta_{-i})$.

\end{theorem}
\Proof Let $j\in\{1,\LL,n\}\setminus\{i\}$ and $\theta'_i$ be such that
$f(\theta'_i, \theta_{-i}) = f(\theta_i, \theta_{-i})=\bar{l}$. Then
\[
t_j(\theta'_i,\theta_{-i}):=\sum_{k\not=i,j}\theta_{k\bar{l}}+\theta'_{i\bar{l}}- 
\max_{k \in \{1, \LL, m\}} (\sum_{k\not=i,j}\theta_{kl}+\theta'_{il}) \leq 0.
\]
If $t_j(s_i(\theta_1,\LL,\theta_i),\theta_{-i})=0$, the result follows immediately. Hence, assume $t_j(s_i(\theta_1,\LL,\theta_i),\theta_{-i})<0$ and let $\tilde{l}=
\arg\max\{\sum_{k\not=i,j}\theta_{kl}+s_{il}(\theta_1,\LL,\theta_i)|l \in \{1, \LL, m\}\}$. Then
\begin{align*}
t_j(s_i(\theta_1,\LL,\theta_i),\theta_{-i})&=\sum_{k\not=i,j}\theta_{k\bar{l}}+s_{i\bar{l}}(\theta_1,\LL,\theta_i)-(\sum_{k\not=i,j}\theta_{k\tilde{l}}+s_{i\tilde{l}}(\theta_1,\LL,\theta_i))\\
&=\sum_{k\not=i,j}\theta_{k\bar{l}}+r_{i\bar{l}}-(\sum_{k\not=i,j}\theta_{k\tilde{l}}+0)\\
&\geq \sum_{k\not=i,j}\theta_{k\bar{l}}+\theta'_{i\bar{l}}-(\sum_{k\not=i,j}\theta_{k\tilde{l}}+\theta'_{i\tilde{l}})\\
&\geq \sum_{k\not=i,j}\theta_{k\bar{l}}+\theta'_{i\bar{l}}-\max_{l \in \{1, \LL, m\}} (\sum_{k\not=i,j}\theta_{kl}+\theta'_{il}) \\
&=t_j(\theta'_i,\theta_{-i}).
\end{align*}

\HB \VV

Let us illustrate now the above two theorems with the following example.

\begin{example} \label{exa:4}
  
  Suppose that it has to be decided which project out of two is going to be realized. Moreover, the decision is going to take place depending on the valuations of three players, A, B, and C, whose types (values) and type spaces are summarized in Table
  \ref{tab:4}.\\

\begin{table}[htbp]

\begin{center}
\begin{tabular}{|c|c|c|c|c|c|}
\hline
player & value 1 & value 2 & set of types ($\Theta_i$) & Clarke tax & utility ($u_i$) \\\hline
A & $6$  & $9$   & $[0,9]\times[0,10]$  & $1$            & $8$\\\hline
B & $12$ & $1$   & $[0,12]\times[0,2]$  & $0$  & $0$\\\hline
C & $30$ & $40$  & $[0,34]\times[0,40]$ & $8$            & $32$ \\\hline
\end{tabular}\vspace{0.25cm}
\end{center}
  \caption{Clarke taxes: project 2 takes place.}
  \label{tab:4}
\end{table}

Consider now the situation in which the players submit their values sequentially
and each of them follows strategy  $s_{i}(\cdot)$.
There are three possible cases. 

\begin{itemize}
\item Player A is the last player. 
  
\NI
According to strategy $s_{i}(\cdot)$ players B and C will submit their
true values, since it is not known which project will be chosen before
A's submission. Therefore in the definition of $s_{i}(\cdot)$ the
first alternative holds for each of them.  However, player A will
submit $(0,10)$ since he knows that project 2 will take place and
therefore for him the second alternative holds. The tax of player A remains 1,
that of player B remains 0, but the tax of player C gets modified becoming $1$ instead
of $8$.

\item Player B is the last player. 
  
\NI
In this situation, the second player always knows which project will
be chosen. Hence player A will submit $(0,10)$ when he is second and player C will
submit $(0,40)$ when he is second. As a result, all taxes become 0.

 \item Player C is the last player. 
   
\NI
This situation is similar to the first one. Here only player C knows
that project 2 will be chosen before his submission and can deviate
from truth-telling. According to the $s_{i}(\cdot)$ strategy
player C will submit $(0,40)$ while players
A and B will submit their true values. This does not modify the taxes of
players B and C (which remain $0$ and $8$) but player's A tax becomes $0$.

\end{itemize}

The situation is summarized in Table \ref{tab:4a}.

\begin{table}[ht]

\begin{center}
\begin{tabular}{|c|c|c|c|}
\hline
ordering &  $t_A$         &  $t_B$  & $t_C$ \\\hline
A B C    & $0$ & $0$ & $8$  \\\hline
A C B    & $0$ & $0$ & $0$  \\\hline
B A C    & $0$ & $0$ & $8$  \\\hline
B C A    & $1$ & $0$ & $1$  \\\hline
C A B    & $0$ & $0$ & $0$  \\\hline
C B A    & $1$ & $0$ & $1$  \\\hline
\end{tabular}\vspace{0.25cm}
\end{center}
  \caption{Clarke taxes in the sequential cases of Example~\ref{exa:4}.}
  \label{tab:4a}
\end{table}
\HB
\end{example}

 \bibliography{/ufs/apt/bib/e,/ufs/apt/bib/apt}
 \bibliographystyle{handbk}

\end{document}
%%CHANGE TO NEW SETTING

% \begin{note}
%   The decision rule $f$ is strategy-proof iff for all $i \in \C{1,\LL,n}$
%   the identity function $I_i: \Theta_i \myra \Theta_i$ is a
%   dominant strategy for player $i$ in the corresponding pre-Bayesian
%   game.
% \end{note}
%
% \Proof
%
% %$
% %p_i(s_1, \LL, s_n, \theta) = v_i(f(s_1, \LL, s_n)}, \theta_i).
% %$
% %\II
%
% \NI
% Choose arbitrary $\theta \in \Theta$ and $\theta'_i \in \Theta_i$.
% We have
% \[
% p_i(I_i(\theta_i), \theta_{-i}, \theta_i) = v_i(f(\theta_i, \theta_{-i}), \theta_i)
% \]
% and
% \[
% p_i(\theta'_i, \theta_{-i}, \theta_i) = v_i(f(\theta'_i, \theta_{-i}), \theta_i).
% \]
% So
% \[
% \mbox{$v_i(f(\theta_i, \theta_{-i}), \theta_i) \geq v_i(f(\theta'_i,
% \theta_{-i}), \theta_i)$ iff $p_i(I_i(\theta_i), \theta_{-i}, \theta_i)
% \geq p_i(\theta'_i, \theta_{-i}, \theta_i)$},
% \]
% which concludes the proof.
% \HB

\section{Set up}

Recall that a \oldbfe{preorder} $\succeq$ on a set $A$ is a transitive and
reflexive relation on $A$.  We denote by $\succ$ the irreflexive
relation corresponding to $\succeq$. That is, $a \succ b$ iff $a
\succeq b$ and not $b \succeq a$.

Given a preorder $\succeq$ on $A$, a subset $B$ of $A$ and $a \in B$
we say that $a$ is a \oldbfe{$\succeq$-maximal element in $B$} if for no $b
\in B$ we have $b \succ a$ and
a \oldbfe{strict $\succeq$-maximal element in $B$} if for no $b
\in B$ we have $b \succeq a$.

We begin with a more general definition of a strategic game, which
is now defined as
\[
G := (S_1, \LL, S_n, \succeq_1, \LL, \succeq_n),
\]
where for each $i \in \C{1, \LL, n}$

\begin{itemize}
\item $S_i$ is the non-empty set of \oldbfe{strategies}
available to player $i$,

\item $\succeq_i$ is a preorder
over the set of joint profiles
$S := S_1 \times \LL \times S_n$ that represents
the \oldbfe{preference} of player $i$.
\end{itemize}

We call a strategy profile $s$ a \oldbfe{Nash equilibrium of} $G$ if
for all $i \in \C{1, \LL, n}$ we have that $s$ is a
$\succeq_i$-maximal element in the set $\C{(s'_i, s_{-i}) \mid s'_i
  \in S_i}$.  Further, we call $s$ a \oldbfe{strict Nash equilibrium
    of} $G$ if for all $i \in \C{1, \LL, n}$ we have that $s$ is a
  strict $\succeq_i$-maximal element in the set $\C{(s'_i, s_{-i})
    \mid s'_i \in S_i}$.

Let $N := \C{1,\LL,n}$.
We now consider a situation in which the players choose their
strategies \emph{sequentially} starting with player 1 and ending with
player $n$.

We assume that each player $i$ is a $\succeq_i$-maximizer.  We
formalize this assumption in the sequential setting by defining for
each $i \in N$ a function
\[
\emph{opt}_i: S_1 \times \LL \times S_{i-1} \myra {\cal P}(S_i)
\]
that associates with each sequence $s_1, \LL, s_{i-1}$ of strategies
of players $1, \LL, i-1$
the set of `optimal' strategies of player $i$.
In particular $\emph{opt}_1 \sse S_1$.

The functions $\emph{opt}_i$ are defined by induction, starting with
player $n$:
\[
opt_n(s_1, \LL, s_{n-1}) := \C{s_n \in S_n \mid (s_1, \LL, s_n) \mbox{ is an $\succeq_n$-maximal element in $\emph{Opt}_n(s_1, \LL, s_{n-1})$}},
\]
\[
\emph{opt}_i(s_1, \LL, s_{i-1}) := \C{s_i \in S_i \mid (s_1, \LL, s_n) \mbox{ is an $\succeq_i$-maximal element in $\emph{Opt}_i(s_1, \LL, s_{i-1})$}},
\]
where
\[
\emph{Opt}_i(s_1, \LL, s_{i-1}) := \C{(s_1, \LL, s_n) \mid s_i \in S_i,  \fa j \in \C{i+1, \LL, n} \: s_j \in \emph{opt}_j(s_1, \LL, s_{j-1})}.
\]

Note that $\emph{Opt}_n(s_1, \LL, s_{n-1}) =
\C{s_1} \times \LL \times \C{s_{n-1}} \times S_n$.

Let now
\[
\emph{Opt} := \C{(s_1, \LL, s_n) \mid \fa j \in \C{1, \LL, n} \: s_j \in \emph{opt}_j(s_1, \LL, s_{j-1})}.
\]

This set consists of the strategy profiles that will be selected
by the players.
\section{Specific preferences}

The above general set-up leans itself for various specializations that we now describe.

\paragraph{Categorizing opponents}

We start by assuming that the preference of each
player $i$ is represented by means of a \oldbfe{payoff function} $p_i
: S_1 \times \LL \times S_n \myra \cal{R}, $ where $\cal{R}$ is the
set of real numbers.

Then we associate with each player $i$ two disjoint sets of
players not including $i$: his associates, $A(i)$ and his adversaries,
$E(i)$. Thus each player $i$ divides the set of his opponents into
three categories: associates, adversaries and others, denoted by $O(i)$.

So for each $i \in N$ we have $\C{i} \cup A(i) \cup E(i) \cup O(i) = N$.
We assume that each player

\begin{itemize}
\item wants to maximize his payoff,
\item wants to help his associates,
\item wants to punish his adversaries,
\item is indifferent to the remaining players,
\end{itemize}
and that he prioritizes these matters in that order.

This can be formalized by first introducing for each player $i$ the
following partial order $\succ_i$ on the strategy profiles from
$S$:

\begin{tabbing}
\quad $s \succ_i s'$ iff \= $p_i(s) > p_i(s')$ \=or \kill
\quad $s \succ_i s'$ iff \> $p_i(s) > p_i(s')$ or \\
                         \> $p_i(s) = p_i(s')$ and $\fa j \in A(i)\: p_j(s) \geq p_j(s')$ and $\te j \in A(i)\: p_j(s) > p_j(s')$ or \\
                         \> $p_i(s) = p_i(s')$ and $\fa j \in A(i)\: p_j(s) = p_j(s')$ and $\fa j \in E(i) \: p_j(s) \leq p_j(s')$\\
               \> \>and $\te j \in E(i) \: p_j(s) < p_j(s')$,
\end{tabbing}
and then defining the preorder $\succeq_i$ by
\[
\mbox{$s \succeq_i s'$ iff $s \succ_i s'$ or ($p_i(s) = p_i(s')$ and $\fa j \in A(i)\: p_j(s) = p_j(s')$ and $\fa j \in E(i) \: p_j(s) = p_j(s')$).}
\]
Note that in each preorder $\succeq_i$ several strategy profiles
can be maximal elements. When both $s \succeq_i s'$ and $s' \succeq_i s$ we write
$s \sim_i s'$.

The above formalization presupposes a complete information, i.e.,
we assume that each player $i$ is fully informed
about the way every other player divides the set of his opponents,
i.e., each player $i$ knows the sets $A(j)$ and $E(j)$ for $j \neq i$.

\paragraph{Formalizing attitudes}

The definition of the set \emph{Opt} of profiles leaves open the possibility that
players may face suboptimal choices by opponents who are indifferent to them.
For example in the case of a game
with two players (a Stackelberg competition)
the second player can have a number of best replies which yield different
payoffs to the first player.

We can incorporate into the above set-up the \emph{optimistic attitude}
according to which player $i$, when facing a choice,
will maximize the payoffs of the players he is indifferent to.
This is achieved by modifying each preorder $\succeq_i$ to the preorder
$\succeq^{o}_i$ defined by:
\[
\mbox{s $\succeq^{o}_i s'$ iff $s \succeq_i s'$ or ($s \sim_i s'$ and $\fa j \in O(i) \: p_j(s) \geq p_j(s')$).}
\]

The corresponding \emph{pessimistic attitude}
is formalized by means of the preorder
$\succeq^{p}_i$ defined by:
\[
\mbox{s $\succeq^{p}_i s'$ iff $s \succeq_i s'$ or ($s \sim_i s'$ and $\fa j \in O(i) \:  p_j(s') \geq p_j(s)$).}
\]

Both attitudes can be alternatively formalized (CHECK) by modifying
the set of associates or adversaries.  The optimistic attitude is obtained
by using for player $i$, $A(i) \cup O(i)$ as the set of
associates and the pessimistic attitude by using for each player $i$,
$E(i) \cup O(i)$ as the set of adversaries.

Finally, we can consider the \emph{social attitude}, according to
which player $i$, when facing a choice, will take a decision that
maximizes the social welfare of the players he is indifferent to.
This is formalized by
means of the preorder $\succeq^{s}_i$ defined by:
\[
\mbox{s $\succeq^{s}_i s'$ iff $s \succeq_i s'$ or ($s \sim_i s'$ and $\sum_{j \in O(i)} p_j(s) \geq \sum_{j \in O(i)} p_j(s')$).}
\]

\paragraph{Incomplete information}

We model incomplete information by assuming that each player $i$ knows
the sets $A(i)$ and $E(i)$ and knows how each player $j \neq i$ views
him (i.e., player $i$ knows to which set $A(j), E(j)$ or $O(j)$ he
belongs) but he does not know the sets $A(j)$ and $E(j)$ for $j \neq
i$.

\paragraph{Beliefs}

In this version we assume that each player $i$ knows the sets $A(i)$
and $E(i)$ and knows how each player $j \neq i$ views him but only has
a belief about the way each player $j \neq i$ divides the set of
players $N \setminus \C{i,j}$.

Such a belief of player $i$ about player $j$ is a sequence of the
probability distributions $(\pi_{i,j,k})_{k \neq i,j}$ such that for
each $k \in N \setminus \C{i,j}$, $\pi_{i,j,k}$ assigns to $k$ the
probability that $k$ is in one of the sets $A(j), E(j)$ or $O(j)$.

To define the set of the strategy profiles that will be selected
by the players we need to accommodate the probability distributions
into the previous definitions.

We start with the relation $\succ_i$ and define its modification
$\succ^j_k$ that represents player's $k$ belief about the relation
$\succ_i$ used by player $i$.

\end{document}

Clearly all dominant strategies for player $i$
yield the same payoff to him, so from player $i$ point of view it does
not matter which strategy he selects. However, the choices made by
each player \emph{can} affect the payoffs of the other players.  This
can be easily seen by considering a simple instance of the above
situation.

\begin{game}{2}{2}
      & $L$    & $R$\\
$T$   &$0,0$   &$5,0$\\
$B$   &$0,5$   &$5,5$
\end{game}

\II

In this game both strategies of player 1 are dominant and also both
strategies of player 2 are dominant. So, in particular $(T,L)$
is a Nash equilibrium, but clearly not Pareto optimal.

If this game is played sequentially, with player 1 starting, then
player 1 can help player 2 by selecting $B$. This secures payoff 5 to
player 2. In turn, player 2 can always help player 1 by selecting $R$
and securing payoff 5 to player 1.  Analogously, player 1 can punish
player 2 by selecting $T$ and player 2 can punish player 1 by
selecting $L$, each time imposing the payoff 0 on the other player.
The Pareto optimal outcome $(B,R)$ is ensured when each player decides
to help the other.

To model such situations we need to modify the underlying preferences of the players.

Clearly all dominant strategies for player $i$
yield the same payoff to him, so from player $i$ point of view it does
not matter which strategy he selects. However, the choices made by
each player \emph{can} affect the payoffs of the other players.  This
can be easily seen by considering a simple instance of the above
situation.

\begin{game}{2}{2}
      & $L$    & $R$\\
$T$   &$0,0$   &$5,0$\\
$B$   &$0,5$   &$5,5$
\end{game}

\II

In this game both strategies of player 1 are dominant and also both
strategies of player 2 are dominant. So, in particular $(T,L)$
is a Nash equilibrium, but clearly not Pareto optimal.

If this game is played sequentially, with player 1 starting, then
player 1 can help player 2 by selecting $B$. This secures payoff 5 to
player 2. In turn, player 2 can always help player 1 by selecting $R$
and securing payoff 5 to player 1.  Analogously, player 1 can punish
player 2 by selecting $T$ and player 2 can punish player 1 by
selecting $L$, each time imposing the payoff 0 on the other player.
The Pareto optimal outcome $(B,R)$ is ensured when each player decides
to help the other.

To model such situations we need to modify the underlying preferences of the players.

---------------

\end{document}

%% file: ecmds.tex
\newcommand{\oldbfe}[1]{\begin{bfseries}\emph{#1}\end{bfseries}}

\newcommand{\myra}{\mbox{$\:\rightarrow\:$}}

\newcommand{\sse}{\mbox{$\:\subseteq\:$}}

\newcommand{\fa}{\mbox{$\forall$}}
\newcommand{\te}{\mbox{$\exists$}}

\newcommand{\LL}{\mbox{$\ldots$}}

\newcommand{\C}[1]{\mbox{$\{{#1}\}$}}           % curly braces

\newcommand{\NI}{\noindent}
\newcommand{\HB}{\hfill{$\Box$}}
\newcommand{\VV}{\vspace{5 mm}}

\newcommand{\II}{\vspace{2 mm}}

\newcommand{\szkew}[1]{\relax \setbox0=\hbox{\kern -24pt $\displaystyle#1$\kern 0pt }%
\box0}
{\catcode`\@=11 \global\let\ifjusthvtest@=\iffalse}
\newcounter{oldmycaption}

%% file: header.tex
\def\smallromani{\renewcommand{\theenumi}{\roman{enumi}}
\renewcommand{\labelenumi}{(\theenumi)}}

\newcommand{\Proof}{\NI
                    {\bf Proof.}\ }
\newtheorem{theorem}{Theorem}[section]
\newtheorem{defined}[theorem]{Definition}

\newtheorem{exa}[theorem]{Example}
\newenvironment{example}{\begin{exa} \rm}{\end{exa}}

\newtheorem{lemma}[theorem]{Lemma}

\newtheorem{note}[theorem]{Note}

\newtheorem{exe}{Exercise}

\newtheorem{pro}{Problem}

\newcounter{symbol}
\newcommand{\indexsyma}[1]%
{\stepcounter{symbol}\index{zzz1 \thesymbol @\protect#1}}
\newcommand{\indexsymb}[1]%
{\stepcounter{symbol}\index{zzz2 \thesymbol @\protect#1}}
\newcommand{\indexsymc}[1]%
{\stepcounter{symbol}\index{zzz3 \thesymbol @\protect#1}}
\newcommand{\indexsymd}[1]%
{\stepcounter{symbol}\index{zzz4 \thesymbol @\protect#1}}
\newcommand{\indexsyme}[1]%
{\stepcounter{symbol}\index{zzz5 \thesymbol @\protect#1}}

%% file: layout.tex
\normalbaselines

%% file: ae07.bbl
\begin{thebibliography}{1999}

\bibitem[Ashlagi, Monderer and Tennenholtz:\nameindex{Ashlagi,
  I.}\nameindex{Monderer, D.}\nameindex{Tennenholtz, M.}:2006]{AMT06}
{\sc I.~Ashlagi\nameindex{Ashlagi, I.}, D.~Monderer\nameindex{Monderer, D.},
  and M.~Tennenholtz\nameindex{Tennenholtz, M.}}, Resource selection games with
  unknown number of players, in: {\em AAMAS '06: Proceedings of the Fifth
  International Joint Conference on Autonomous Agents and Multiagent Systems},
  ACM Press, New York, NY, USA, pp.~819--825.

\bibitem[Bowles:\nameindex{Bowles, S.}:2004]{Bow04}
{\sc S.~Bowles\nameindex{Bowles, S.}}, {\em Microeconomics: Behavior,
  Institutions, and Evolution}, Princeton University Press, Princeton.

\bibitem[Cavallo:\nameindex{Cavallo, R.}:2006]{Cav06}
{\sc R.~Cavallo\nameindex{Cavallo, R.}}, Optimal decision-making with minimal
  waste: Strategyproof redistribution of {VCG} payments, in: {\em AAMAS '06:
  Proceedings of the Fifth International Joint Conference on Autonomous Agents
  and Multiagent Systems}, ACM Press, New York, NY, USA, pp.~882--889.

\bibitem[Dekel and Piccione:\nameindex{Dekel, E.}\nameindex{Piccione,
  M.}:2000]{DP00}
{\sc E.~Dekel\nameindex{Dekel, E.} and M.~Piccione\nameindex{Piccione, M.}},
  Sequential voting procedures in symmetric binary elections, {\em Journal of
  Political Economy}, 108, pp.~34--55.

\bibitem[Jackson:\nameindex{Jackson, M.}:2003]{Jac03}
{\sc M.~Jackson\nameindex{Jackson, M.}}, Mechanism theory, in: {\em
  Encyclopedia of Life Support Systems}, U.~Derigs\nameindex{Derigs, U.}, ed.,
  EOLSS Publishers, Oxford, UK.

\bibitem[Moore and Repullo:\nameindex{Moore, J.}\nameindex{Repullo,
  R.}:1988]{MR88}
{\sc J.~Moore\nameindex{Moore, J.} and R.~Repullo\nameindex{Repullo, R.}},
  Subgame perfect implementation, {\em Econometrica}, 56, pp.~1191--1220.

\bibitem[Parkes and Shneidman:\nameindex{Parkes, D.~C.}\nameindex{Shneidman,
  J.}:2004]{PS04}
{\sc D.~C. Parkes\nameindex{Parkes, D.~C.} and
  J.~Shneidman\nameindex{Shneidman, J.}}, Distributed implementations of
  {Vickrey-Clarke-Groves} mechanisms, in: {\em Proc. 3rd Int. Joint Conf. on
  Autonomous Agents and Multi Agent Systems}, pp.~261--268.

\bibitem[Varian:\nameindex{Varian, H.}:1994]{Var94}
{\sc H.~Varian\nameindex{Varian, H.}}, Sequential provision of public goods,
  {\em Journal of Public Economics}, 53, p.~165~186.

\end{thebibliography}
